\documentclass[12pt]{article}

\usepackage{amsmath}
\usepackage{amsfonts}
\usepackage{latexsym}
\usepackage{mathrsfs}

\def\C{{\mathscr C}}
\newcommand{\N}{{\mathscr N}}
\newcommand{\B}{{\mathscr B}}
\newcommand{\Bl}{{\mathscr B}_{\rm loc}}
\newcommand{\Blb}{\bar {\mathscr B}_{\rm loc}}

\newcommand{\e}{{\rm e}}   
\newcommand{\I}{{\rm i}}

\def\l{\langle}
\def\r{\rangle}
\def\pr{\partial}

\newcommand{\half}{{\textstyle \frac{1}{2}}}
\def\quar{{\textstyle \frac{1}{4}}}

\newcommand{\dx}{\!\!{\rm d}^4x\,\,}
\newcommand{\dS}{\!\!{\rm d}^6z\,}
\newcommand{\dSb}{\!\!{\rm d}^6\bar z\,}
\newcommand{\dV}{\!\!{\rm d}^8z\,}

\newcommand{\Leff}{L_{\rm eff}}
\newcommand{\Leffb}{\bar L_{\rm eff}}

\newcommand{\Geff}{\Gamma_{\rm eff}}
\newcommand{\Gcl}{\Gamma_{\rm cl}}
\newcommand{\Lkin}{L_{\rm kin}}
\newcommand{\Lxi}{L_{\xi}}
\newcommand{\Lxib}{L_{\xi}'{}}
\newcommand{\ukin}{u_{\rm kin}}

\newcommand{\uxi}{u_{\xi}}
\newcommand{\uxib}{{u_{\xi}'}}

\newcommand{\al}{\alpha}
\newcommand{\da}{{\dot\alpha}}
\newcommand{\be}{\beta}
\newcommand{\db}{{\dot\beta}}
\newcommand{\eps}{\varepsilon}

\newcommand{\tfr}[2]{{\textstyle \frac{#1}{#2}}}

\newcommand{\fdq}[2]{\frac{\delta #1}{\delta #2}}

\begin{document}

\thispagestyle{empty}

\vspace*{0.1cm}
\begin{center}
{\LARGE \bf Supercurrent and Local Coupling\\[2mm] in the Wess-Zumino
  Model}

\vspace{1cm}

Elisabeth Kraus${}^a$,
Christian Rupp${}^b$, 
Klaus Sibold${}^c$,
\\[0.7cm]
${}^a$ Physikalisches Institut der Universit\"at Bonn,\\
Nu\ss allee 12, D - 53115 Bonn, Germany\\
Email: kraus@th.physik.uni-bonn.de
\\[5mm]
${}^b$ Institut f{\"u}r Theoretische Physik, Universit{\"a}t Bern,\\
Sidlerstrasse 5, CH - 3012 Bern, Switzerland\\
Email: rupp@itp.unibe.ch
\\[5mm]
${}^c$ Institut f{\"u}r Theoretische Physik, Universit{\"a}t Leipzig,\\
Augustusplatz 10/11, D - 04109 Leipzig, Germany\\
Email: Klaus.Sibold@itp.uni-leipzig.de\\

\end{center}

\vspace{0.5cm}

\begin{center}
\parbox{12cm}{
\centerline{\small \bf Abstract}
\small \noindent 
We study the Wess-Zumino model with the coupling extended to a chiral
superfield. In order to incorporate the renormalization effects a
further external {\sl real} field  has to be introduced. It is then
possible to derive a Callan-Symanzik equation and to prove renormalizability.
By constructing the supercurrent in this context the whole machinery
for describing the superconformal symmetries becomes available.
The presence of the external fields allows also to define multiple
insertions of all relevant composite operators. Interesting relations
to the curved superspace treatment show up.
}
\end{center}

\vspace*{10mm}
\begin{tabbing}
\> Wess-Zumino Model, Local Coupling
\end{tabbing}
\newpage

\section{Introduction}
\setcounter{equation}{0}

The non-renormalization of chiral vertices has been a key issue
in supersymmetric theories from the very first moment when
one looked at the renormalization problem.
At the beginning it was seen in explicit component calculations
 \cite{WZ}
and then automatically realized when
performing perturbation theory in terms of superfields \cite{FuLa, FePi,GSR79}.
The puzzling point has always been that the supersymmetry Ward identities
(WI) did not require them to hold, but the supersymmetric structure
of the {\sl integrands} -- maintained when working with supergraphs --
ensured them. 

A deeper understanding has been achieved only recently
\cite{FK} using two ingredients: (1) rendering the coupling local
causes flow of external momentum through every internal line of
a diagram and thus makes the integrand accessible to operations
``from the outside''; it also opens the way to the lowest $\theta$ component
of a vertex which has lower dimension than the highest one. (2)
observing and exploiting the fact that an integrated vertex is always
a susy variation and thus carries momentum factors.
Quite a number of results have been obtained meanwhile which
clearly show that this new insight is fruitful
\cite{KRST01,KR01,KR01anom}. This refers in
particular to susy gauge theories formulated in the Wess-Zumino
gauge  where these non-renormalization theorems would not be
available in any other way than by falling over them in
explicit calculations. 

It is then obvious that one should
study the effect of local couplings also in the context
of linear realization of supersymmetry in terms of superfields 
and exploit the new tool. Since in the past it has become
evident at many instances that the {\sl supercurrent} is the
carrier of the most important information on local structure
in susy models we formulate in the present paper the
Wess-Zumino model as an example of a chiral model with
local coupling. We derive the supercurrent in presence of
a local coupling and thus pave the way for the analysis
of the whole superconformal properties of the model. 
In particular we derive a local and an integrated
Callan-Symanzik equation which completes the renormalizability proof
in \cite{FK} when the coupling is local.

The paper is organized as follows. In Sect. 2 we set up global
WI's for two different R-symmetries and their difference which
plays a decisive role in the subsequent construction. For it
gives raise to a powerful local gauge WI which in Sect. 3
is used to define the multiple insertions of the composite
operator $A\bar A$ which contains an axial current as one
of its components. It is essentially this operator which is
responsible for all renormalization effects in the model.
In Sect. 4 we derive the local and integrated Callan-Symanzik equation
via the connection of the dilatations with scaling of
parameters carrying mass dimension. In Sect. 5 we
derive as an application some properties of double
insertions showing thereby the usefulness of the local
coupling and the external field introduced for handling
the product $A\bar A$. The connection to the curved superspace
treatment found in particular there clearly indicates that our results are
as scheme independent as they can be when one is
constructing currents and the like {\sl explicitly}. 
In the conclusions we discuss
our findings and give an outlook to further applications
of the local coupling.

\section{Global Ward Identities} \label{sec:GlobalWI}
\setcounter{equation}{0}

Introducing a local coupling in the Wess-Zumino model ($A$: chiral
superfield)
\begin{equation}
\Gcl = \tfr{1}{16} \int \dV A \bar A + \int \dS \left(\tfr{m}{8} A^2 +
  \tfr{\lambda}{48} A^3 \right) 
 + \int \dSb \left(\tfr{m}{8} \bar A^2 +   \tfr{\lambda}{48} \bar A^3 \right) 
\end{equation}
means promoting the real coupling constant $\lambda$ to chiral
external fields
$\Lambda$, $\bar\Lambda$ such that one constructs Feynman diagrams from
\begin{align}
\Geff &= \tfr{1}{16} \int \dV \sum_{n=0}^\infty z^{(n)} \hbar^n
(\Lambda\Bar\Lambda)^n A \bar A \nonumber  \\
&\quad + \int \dS \left( \tfr{m}{8} A^2 + \tfr{1}{48} \Lambda A^3 \right) 
 + \int \dSb \left(\tfr{m}{8} \bar A^2 +   \tfr{1}{48} \bar \Lambda \bar A^3
 \right) 
\nonumber \\
&\quad + \tfr{1}{8} \int \dV \sum_{n=1}^\infty \xi^{(n)} \hbar^{n+1}
(\Lambda \bar\Lambda)^n \left( (\Lambda A)^2 + (\bar\Lambda \bar A)^2
\right) \,.
\end{align}
Due to the space-time dependence of $\Lambda$ and $\Bar\Lambda$
every vertex in a diagram receives external momentum in accordance
with the standard Feynman rules.  

The usage of the above $\Geff$ refers to renormalization
in the BPHZ scheme, which has been shown to be a
supersymmetric invariant scheme. Hence supersymmetric Ward identities
are maintained in the construction and superfield expressions can be
used throughout the paper.
 
The terms going with $z^{(n)}, z^{(0)} = 1$ describe the kinetic term
and terms which reduce to its counterterms in the limit of constant 
coupling. Mass and interaction terms have by prescription no
counterterms. This implies 
the non-renormalization of chiral vertices if renormalization is possible 
with such a choice of $\Geff$. The terms going with $\xi^{(n)}$ represent
power counting admissible counterterms which vanish in the adiabatic limit.
We have omitted possible terms which are linear in the field $A$ because
they do not play a role in what follows and the omission is consistent
with higher orders. 
Powers  of $\Lambda$ and $\bar\Lambda$ are assigned together with orders in
$\hbar$ in such a way that a $R'$-symmetry \cite{SEI93,FK} is maintained
naively
\begin{equation}
W^{R'} \Gamma \equiv \left( -\I \int \dS \I( -1+\theta^\al\pr_\al) A \fdq{}{A}
+ \I( 1+\theta^\al\pr_\al) \Lambda
\fdq{}{\Lambda} - c.c. \right) \Gamma =0
\end{equation}
The validity of this WI guarantees already the non-renormalization of
chiral vertices.

An important member of the superconformal symmetry is yet another
$R$-symmetry:  there the fields have the so called conformal weights.
For constant coupling it is {\sl always} only softly broken. For local
coupling it is again only softly broken 
\begin{multline}
W^{R} \Gamma \equiv \left( -\I \int \dS \I( -\tfr{2}{3}+\theta^\al\pr_\al)
  A \fdq{}{A} + \I\theta^\al\pr_\al \Lambda \fdq{}{\Lambda} - c.c. \right)
  \Gamma \\
= \tfr{m}{12} \left[ \int \dS A^2 - \int \dSb \bar A^2 \right]_3 \cdot
  \Gamma,
\label{RWI}
\end{multline}
if the $\xi^{(n)}$ are appropriately chosen, as will be shown more explicitly
below.  The coefficients $z^{(n)}$ can be fixed by the usual normalization
condition prescribing the wave function renormalization in the flat limit.
It will turn out to be useful to consider the difference of the above two
$R$-symmetries which is a global U(1) commuting with supersymmetry:
 \begin{align}
W^3 \Gamma &\equiv (W^{R'}-W^{R})\Gamma  = \left( \int \dS \left(-\tfr{1}{3} A
  \fdq{}{A} + \Lambda \fdq{}{\Lambda}\right) -c.c. \right) \Gamma \label{W3def}\\
W^3\Gamma &= -\tfr{m}{12} \left[ \int \dS A^2 - \int \dSb \bar A^2
  \right]_3 \cdot \Gamma\,.
\end{align}

One of the aims in the present paper is the derivation of the
Callan-Symanzik equation (CS) which in the adiabatic limit
($\Lambda=\bar\Lambda=\lambda$) is known \cite{CPS,PSbook} to have the
form
\begin{equation}
\left( m\pr_m + 2 \kappa^2 \pr_{\kappa^2} + \beta \pr_\lambda -\gamma \N
\right)\Gamma = \alpha \tfr{m}{4} \left[ \int \dS A^2 + \int \dSb \bar
  A^2\right]_3\cdot \Gamma
\label{CSstandard}
\end{equation}
with
\begin{equation}
\beta=3\gamma\,, \qquad \alpha-1 = 2\gamma\,.
\end{equation}
$\gamma$ denotes the anomalous dimension which will be given more
explicitly below. $\kappa^2$ is the normalization point where the wave
function counterterm is fixed. As long as the coupling $\Lambda$ is local
one cannot express the effect of $m\pr_m + 2\kappa^2 \pr_{\kappa^2}$ on
$\Gamma$ by other differential operators and a soft mass insertion. Hence
for local coupling
one has to introduce another external field $L$ which is a real superfield of
dimension and R-weight zero and couples accordingly;
  in particular it
couples to the product $A \bar A$ and allows to absorb the hard breaking of the CS equation into a field operator (see also \cite{KRST01,AGLR98}).  
Since by dimensional analysis 
\begin{equation}
\left( m\pr_m + 2\kappa^2 \pr_{\kappa^2} \right) \Gamma = -\I W^D \Gamma\,,
\end{equation}
where
\begin{equation}
W^D \equiv -\I \int \dx \sum_\phi (d(\phi) + x \pr_x) \phi \fdq{}{\phi}
\end{equation}
denotes the dilatations and furthermore a supersymmetric extension of the
latter can be constructed in a systematic fashion \cite{CPS,PSbook} we go
over to local Ward identities.

\section{Local Ward identities}
\setcounter{equation}{0}

All subsequent  considerations will be based on the following $\Geff$.
\begin{align}
\Geff &= \tfr{1}{16}\int \dV  \sum_{n,m=0}^\infty z^{(n,m)} \hbar^n (\Lambda
  \bar \Lambda)^n L^m A  \bar A \nonumber \\
& \quad
+ \left( \int \dS \left( \tfr{m}{8} A^2 +\tfr{1}{48} \Lambda A^3  \right) +
  c.c. \right) 
 \nonumber \\
& \quad
+ \tfr{1}{8}
\int \dV  \left( \sum_{n,m=0}^\infty \xi^{(n,m)} \hbar^{n+1} 
(\Lambda \bar \Lambda)^{n} L^m \left( (\Lambda A) ^2 + (\bar \Lambda \bar A)
  ^2  \right)  \right) \label{Geff}
\end{align}
The systematic procedure for constructing all currents and WI's of the
superconformal transformations alluded to above (chapts.\ 6 \& 16 in \cite{PSbook})
starts with contact terms
\begin{align}
w_\al^{\rm conf} &= - \tfr{2}{3} D_\al \left(A \fdq{}{A} \right) 
+  2 D_\al A \fdq{}{A} 
+  2 D_\al\Lambda \fdq{}{\Lambda} \nonumber\\
& \quad - 2\bar D^2  D_\al L \fdq{}{L}  + 2D_\al  L \bar D^2 \fdq{}{L}
\end{align}
which give rise e.g.\ via
\begin{equation}
\hat W_R^{\rm conf} \equiv \int \dx \left( D^\al w_\al^{\rm conf} - \bar
  D_\da \bar w^{{\rm conf}\, \da} \right)
\end{equation}
to the conformal $R$-symmetry contact terms and those of supersymmetry and
translations
\begin{align}
\hat W_R^{\rm conf} &= W_R^{\rm conf} + \theta^\al W_\al^{\rm susy} + \bar
\theta_\da \bar W^{{\rm susy} \, \da}
+ \theta \sigma^\nu \bar\theta W_\nu^{\rm P} \\
W_\al^{\rm susy} &\equiv -\I \int \dS \sum_\phi \delta_\al^{\rm susy} \phi
\fdq{}{\phi} \\
W_\nu^{\rm P} &\equiv  -\I \int \dS \sum_\phi \pr_\nu \phi \fdq{}{\phi} 
\qquad \qquad  (\phi  \in \{ A, \bar A, \Lambda, \bar\Lambda, L\} ) 
\end{align}
Those of the $R'$-symmetry read
\begin{align}
w_\al' &= - D_\al \left(A \fdq{}{A} \right) 
+  2 D_\al A \fdq{}{A} 
+  D_\al \left( \Lambda \fdq{}{\Lambda} \right) + 2 D_\al\Lambda \fdq{}{\Lambda} \nonumber\\
& \quad - 2\bar D^2  D_\al L \fdq{}{L}  + 2D_\al  L \bar D^2 \fdq{}{L}
 \,.
\end{align}
The corresponding ones for the ``3'' transformations are in a first step
just the difference
\begin{equation}
w_\al^3 \equiv -\tfr{1}{3} D_\al \left( A \fdq{}{A} \right) + D_\al
\left(\Lambda \fdq{}{\Lambda}\right) \,.
\end{equation}
But noting that they can be obtained by the application of $D_\al$ to more
elementary terms and further identifying those as gauge transformations one
is led to introduce
\begin{align}
  w^3 &\equiv -\tfr{1}{3} A \fdq{}{A} + \Lambda \fdq{}{\Lambda} \\
\tilde w^3 &\equiv w^3 +\tfr{1}{3} \bar D^2 \fdq{}{L} \,.
\label{tildew3}
\end{align}
A quick check on
\begin{equation}
\Gcl = \tfr{1}{16} \int \dV A \e^{L} \bar A 
+ \int \dS \left(\tfr{m}{8} A^2 +
  \tfr{\lambda}{48} A^3 \right) 
 + \int \dSb \left(\tfr{m}{8} \bar A^2 +   \tfr{\lambda}{48} \bar A^3 \right)
\end{equation}
yields
\begin{equation}
\tilde w^3 \Gcl = -\tfr{m}{12} A^2\,, \label{tw3WIclass}
\end{equation}
i.e.\ a local WI where the current is absorbed by the inhomogeneous term of
(\ref{tildew3}) and which is only broken by a mass term.
The contact terms $\tilde w^3$ can be understood as a building block for an
axial transformation, $\tilde w^3- \bar{\tilde w}^3$, but of course also
for all other elements of the superconformal group which makes them
extremely useful.

We shall now apply these WI operators to $\Geff$ and decompose the results
as suggested by the supercurrent treatment \cite{CPS, PSbook}:
\begin{align}
-2 w_\al \Gamma &= \Delta_\al \cdot \Gamma \\
\Delta_\al \cdot  \Gamma &= \bar D^\da V_{\al\da} \cdot \Gamma + 2 D_\al S
\cdot \Gamma - B_\al \cdot \Gamma \, \label{latexm}\\
\left( D^\al w_\al - \bar D_\da \bar w^\da \right) \Gamma &= -\I \pr^\mu
V_\mu \cdot \Gamma + \left( D^2 S - \bar D^2 \bar S \right) \cdot \Gamma \\
\intertext{where}
V_\mu &\equiv \sigma_\mu^{\al\da} V_{\al\da}\\
D^\al B_\al - \bar D_\da \bar B^\da &= 0\,.
\end{align}
The decomposition (\ref{latexm}) into current, $S$-type and $B$-type breaking
represents the most general situation. If a mass term is present the best
one can achieve is a {\sl soft} mass term plus breaking of $S$-type which
means that all superconformal anomalies can be represented in terms of $S$.
The supercurrent contains in this case a current for a softly broken
$R$-symmetry, strictly conserved supersymmetry currents and a conserved
energy-momentum tensor. In the massless case one can have the $B$-type
breaking which implies that {\sl all} components of the supercurrent are
strictly conserved currents and the breaking of superconrformal symmetry is 
described in terms of a real multiplet $B$. In the present case we
let mimic the $R'$-supercurrent the situation with $B$-type breaking
because the mass term is $R'$-invariant. The $S$-type breaking is
represented by the ``conformal'' contact terms, current and breaking,
whereas the ``3'' interpolates between the two and will contain both
types of breaking. Parenthetically we remark that in the curved superspace
treatment this assignment arises automatically.\\
Switching back and forth between $R'$ and $R^{\rm conf}$ and using at
appropriate places the information originating from $\tilde w^3$ we shall
be able to derive first a local and then a global CS equation. 
These calculations become more involved than in the classical approximation
because in higher orders the breaking going with the mass should be soft,
hence one needs first of all a Zimmermann identity relating the
oversubtracted mass term to the minimally subtracted one. It reads
\begin{align}
[m A^2]_3\cdot\Gamma &= [m A^2]_2\cdot\Gamma
\nonumber \\
& \quad + \sum_{n=1,m=0}^\infty \hbar^n \left[ \ukin^{(n,m)} \Lkin^{(n,m)} +
    \uxi^{(n-1,m)} \Lxi^{(n-1,m)} + \uxib^{(n-1,m)} \Lxib^{(n-1,m)}
  \right]_3\cdot\Gamma \nonumber\\
& \quad + \Delta_{\rm triv} \label{Zimmermann}
\end{align}
with
\begin{align}
\Lkin^{(n,m)} &= \bar D^2 \left( (\Lambda \bar \Lambda)^n L^m A \bar A \right) 
 \\
\Lxi^{(n,m)} &= \bar D^2 \left( (\Lambda \bar \Lambda)^n  L^m (\Lambda A)^2
 \right) 
\\
\Lxib^{(n,m)} &= \bar D^2 \left( (\Lambda \bar \Lambda)^n
 L^m (\bar \Lambda \bar   A) ^2 \right)
\end{align}
Here, $\Delta_{\rm triv}$ contains all terms with zero or one propagating
fields. These are trivial in the sense that they cannot contribute to 1PI
loop diagrams in the adiabatic limit and do not affect the discussion of
the non-trivial
breaking terms. In this paper we do not consider $\Delta_{\rm triv}$,
however one should keep in mind that these terms contribute to Ward
identities if tested w.r.t\ less than two dynamical fields.

Our first aim is the local gauge WI
\begin{align}
\tilde w^3 \Gamma &=  \left[ \tilde w^3 \Geff\right]_3\cdot \Gamma\\
\tilde w^3 \Gamma &=  -  \tfr{1}{12}  m \Bigl[ A^2 \Bigr]_2 \cdot
\Gamma \nonumber \\
& \quad +\Bigl[ \tfr{1}{16}\sum_{n,m=0}^\infty
  z^{(n,m)} \hbar^n \left\{(n-\tfr{1}{3}) \Lkin^{(n,m)} +\tfr{m}{3} 
  \Lkin^{(n,m-1)}\right\} \nonumber \\
&\quad\quad\quad -\tfr{1}{12} \sum_{n=1,m=0}^\infty \ukin^{(n,m)}
  \hbar^n \Lkin^{(n,m)} \Bigr]_3\cdot\Gamma \nonumber \\
&\quad + \Bigl[ \tfr{1}{8} \sum_{n,m=0}^\infty \xi^{(n,m)} \hbar^{n+1}
  \left\{ (n+\tfr{4}{3})\Lxi^{(n,m)} +\tfr{m}{3} \Lxi^{(n,m-1)} + n
  \Lxib^{(n,m)}  +\tfr{m}{3} \Lxib^{(n,m-1)} \right\} \nonumber \\
&\quad\quad\quad - \tfr{1}{12} \sum_{n=0,m=0}^\infty
  \hbar^{n+1} (\uxi^{(n,m)} \Lxi^{(n,m)} +  \uxib^{(n,m)} \Lxib^{(n,m)})
  \Bigr]_3\cdot\Gamma 
\label{tw3WI1}
\end{align}
We would like to dispose over the arbitrary coefficients $z^{(n,m)}$ and
$\xi^{(n,m)}$ in such a way that the r.h.s.\ of (\ref{tw3WI1}) reduces to
the soft term. If this is possible we are closest to the classical
situation as expressed by (\ref{tw3WIclass}). 
We have to solve the equations
\begin{align}
\tfr{3}{4} (n-\tfr{1}{3}) z^{(n,m)} + \quar (m+1) z^{(n,m+1)} -
\ukin^{(n,m)}&=0 \label{abs1}\\
\tfr{3}{2}(n+\tfr{4}{3}) \xi^{(n,m)} +\half(m+1) \xi^{(n,m+1)} -\uxi^{(n,m)}&=0 \label{abs2}\\
\tfr{3}{2} n \xi^{(n,m)}  +\half(m+1) \xi^{(n,m+1)}  -\uxib^{(n,m)}&=0 \label{abs3}
\end{align}
(\ref{abs1}) with $z^{(0,0)}=1$ is readily solved by
\begin{align}
z^{(0,m)} &= \frac{1}{m!} \label{zclass}\\
z^{(n,0)} & = \text{arbitrary for } n\ge 1\\
(m+1) z^{(n,m+1)} &= (1-3n) z^{(n,m)}
+4 \ukin^{(n,m)} 
\quad \text{for }n\ge 1, m\ge 0 \label{zsolv}
\end{align}
The difference of (\ref{abs3}) and (\ref{abs2}) determines
\begin{equation}
\xi^{(n,m)} = \half ( \uxi^{(n,m)} - \uxib^{(n,m)} ) \label{xisolve}
\end{equation}
with $\xi^{(0,0)}=0$.

In order to show that  (\ref{abs3}), (\ref{abs2}) hold also separately,
we consider the consistency condition 
\begin{equation}
[\tilde w^3(z), \bar{\tilde w}^3(z')]=0 \,,
\end{equation}
which leads to
\begin{equation}
(n+\tfr{4}{3}) \uxib^{(n,m)} +\tfr{1}{3} (m+1) \uxib^{(n,m+1)} = n
\uxi^{(n,m)} +\tfr{1}{3} (m+1) \uxi^{(n,m+1)} \,.
 \label{consist}
\end{equation}
By inserting (\ref{consist}) into (\ref{xisolve}), one finds that both
(\ref{abs3}) and (\ref{abs2}) are satisfied.

Thus we have
\begin{equation}
\tilde w^3 \Gamma = \Bigl( w^3 +\tfr{1}{3} \bar D^2 \fdq{}{L} \Bigr) \Gamma
=-\tfr{1}{12} m [A^2]_2\cdot
\Gamma + Q^3_{\rm triv}\,, \label{gaugeinv} 
\end{equation}
i.e.\ also on the quantized level we have a gauge theory with abelian gauge
invariance which is only broken a soft mass term.
With (\ref{zclass}) the respective classical action has
the form
\begin{equation}
\Gcl = \tfr{1}{16} \int \dV A \e^{L} \bar A 
+ \int \dS \left(\tfr{m}{8} A^2 +
  \tfr{\lambda}{48} A^3 \right) 
 + \int \dSb \left(\tfr{m}{8} \bar A^2 +   \tfr{\lambda}{48} \bar A^3 \right)
\end{equation}

The result (\ref{gaugeinv}) can be brought very simply into a form where
the machinery of forming moments is applicable and thus the complete
superconformal structure becomes available.
We rewrite (\ref{gaugeinv}) as a trace equation.
\begin{align}
w_\al^3 &\equiv D_\al w^3 = -\tfr{1}{3} D_\al \bar D^2 \fdq{\Gamma}{L}
-\tfr{1}{12} [m D_\al A^2]_{5/2}\cdot\Gamma\\
-2  w_\al^3 \Gamma &=  \bar D^\da V^3_{\al\da}\cdot \Gamma 
+ 2 D_\al S^3 \cdot \Gamma -
B^3_\al \cdot\Gamma
\label{w3al}
\\
V_{\al\da}^3 &\equiv \tfr{4}{3} [D_\al, \bar D_\da] \fdq{\Geff}{L} \\
S^3 &\equiv -\tfr{1}{12} mA^2\\
B_\al^3 &\equiv  2 \bar D^2 D_\al \fdq{\Geff}{L}\\
\fdq{\Geff}{L} &=\tfr{1}{16}\sum_{n=0,m=1}^\infty z^{(n,m)} \hbar^n m L^{m-1}
(\Lambda \bar \Lambda)^n A \bar A \nonumber \\
& \quad + \tfr{1}{8} \sum_{n=0,m=1}^\infty \xi^{(n,m)} \hbar^{n+1} m
L^{m-1} (\Lambda\bar \Lambda)^n (\Lambda^2 A^2 + \bar \Lambda^2 \bar A^2)\end{align}
The next aim is now to derive the trace equation for the conformal contact
terms. Together with (\ref{w3al}) it will contain all information on the
model. 
For the supercurrent generated by the conformal $R$-symmetry we find
\begin{align}
V_{\al\da}^{\rm conf} &= -\tfr{1}{6} \sum_{n,m} z^{(n,m)} \hbar^n \Bigl(
L^m (D_\al (\Lambda^n A) \bar D_\da (\bar \Lambda^n \bar A) - \Lambda^n A D_\al \bar
D_\da (\bar \Lambda^n \bar A) + \bar D_\da D_\al (\Lambda^n A) \bar \Lambda^n \bar A )
\nonumber \\
& \qquad \qquad \qquad \quad \quad
+ D_\al L^m \Lambda^n A \bar D_\da (\bar \Lambda^n \bar A)
 - \bar D_\da L^m D_\al (\Lambda^n
A) \bar \Lambda^n \bar A \nonumber \\
& \qquad \qquad \qquad  \quad \quad
+ \Lambda^n A \bar \Lambda^n \bar A (-m L^{m-1} [D_\al, \bar D_\da] L + m(m-1) L^{m-2}
D_\al L \bar D_\da L ) \Bigr) \nonumber \\
& \quad -\tfr{1}{3} \sum_{n,m} \xi^{(n,m)} \hbar^{n+1} \Bigl(
L^m (D_\al (\Lambda^{n+2} A^2) \bar D_\da (\bar \Lambda^n)
\nonumber \\
& \qquad \qquad \qquad \quad \quad
 - \Lambda^{n+2} A^2 D_\al \bar
D_\da (\bar \Lambda^n) + \bar D_\da D_\al (\Lambda^{n+2} A^2) \bar \Lambda^n)
\nonumber \\
& \qquad \qquad \qquad \quad \quad
+ D_\al L^m \Lambda^{n+2} A^2 \bar D_\da (\bar \Lambda^n) - \bar D_\da L^m D_\al (\Lambda^{n+2}
A^2) \bar \Lambda^n \nonumber \\
& \qquad \qquad \qquad  \quad \quad
+ \Lambda^{n+2} A^2 \bar \Lambda^n  (-m L^{m-1} [D_\al, \bar D_\da] L + m(m-1) L^{m-2}
D_\al L \bar D_\da L ) 
\nonumber \\
& \qquad \qquad \qquad \quad \quad
+ c.c. \Bigr) 
\end{align}
It is instructive to extract the classical approximation
\begin{align}
V_{\al\da}^{\rm conf, class} &= -\tfr{1}{6} \e^L\Bigl(
D_\al A \bar D_\da \bar A - A D_\al \bar D_\da \bar A + \bar D_\da D_\al A
\bar A \nonumber \\
& \qquad \qquad \qquad
+ D_\al L A \bar D_\da \bar A - \bar D_\da L D_\al A \bar A
\nonumber \\
& \qquad \qquad \qquad
-  [D_\al, \bar D_\da] L A\bar A +  D_\al L \bar D_\da L A \bar A \Bigr)
\end{align}
and also to present the limit $L=0$, $\Lambda=\bar \Lambda=\lambda=const$.
\begin{align}
V_{\al\da}^{\rm lim}&= -\tfr{1}{6} z (D_\al A \bar D_\da \bar A -A D_\al
\bar D_\da \bar A + \bar D_\da D_\al A \bar A ) \nonumber \\
& \quad -\tfr{1}{3} \xi \{D_\al, \bar D_\da \} (A^2-\bar A^2)
\end{align}
with
\begin{align}
z &= \sum_n z^{(n,0)} \hbar^n \lambda^{2n}\\
\xi &= \sum_n \xi^{(n,0)} \hbar^{n+1} \lambda^{2n+2} \\
\xi^{(n,0)} &= \half (\uxi^{(n,0)} - \uxib^{(n,0)})
\end{align}
We see that our construction yields a $\xi$ contribution to the
supercurrent precisely as the curved superspace treatment \cite{ERS1}
does. Hence local coupling resolves the decomposition ambiguity $V
\leftrightarrow S$ for these terms as the invariance requirement with
respect to diffeomorphism.          

The other contributions to the trace equation
\begin{equation}
-2 w_\al^{\rm conf} = \bar D^\da V_{\al\da}^{\rm conf} \cdot \Gamma + 2
 D_\al S^{\rm conf} \cdot \Gamma - B_\al^{\rm conf} \cdot \Gamma
\label{traceeq}
\end{equation}
turn out to be
\begin{align}
S^{\rm conf} &= -\tfr{1}{12} m [A^2]_2 \nonumber\\
& \quad -\tfr{1}{12} \left[ \sum_{n,m} \ukin^{(n,m)} \hbar^n \Lkin^{(n,m)}
  + \sum_{n,m} \half (\uxi^{(n,m)} + \uxib^{(n,m)}) \hbar^{n+1}
  (\Lxi^{(n,m)} + \Lxib^{(n,m)}) \right]_3\\
B_\al^{\rm conf} &=0\,.
\end{align}

\section{The CS equation: local and integrated}
\setcounter{equation}{0}

The trace equation (\ref{traceeq}) leads via the moment construction to the
local dilatation WI
\begin{align}
W^D \Gamma &= \hat W^D \Gamma \Bigr|_{\tilde \theta=0}\\
\hat W^D &= \int \dx w^D\\
w^D &= w^P_{\rm trace} + x^\nu w_\nu^P\\
w^P_{\rm trace} &= \tfr{3}{2} \I \left( D^\al w_\al + \bar D_\da \bar w^\da
\right) \\
w^P_\nu &= \tfr{1}{16} \Bigl( - ( D^2 \bar D_\da w_\al + \bar D^2 D_\al
  \bar w_\da) \sigma_\nu^{\al\da} 
- \sigma_\nu^{\be\db} \{ D_\be, \bar D_\db \} (D^\al w_\al - \bar D_\da
\bar w^\da)\nonumber \\
& \qquad  + 8\I \pr_\nu (D^\al w_\al + \bar D_\da \bar w^\da) \Bigr) \\
w^D \Gamma &= \pr^\mu \hat D_\mu \cdot \Gamma -\tfr{3}{2} \I (D^2 S^{\rm
  conf} + \bar D^2 \bar S^{\rm conf}) \cdot \Gamma
\end{align}
A local CS equation is given by the equation
\begin{align}
w_{\rm trace}^P \Gamma \equiv \tfr{3}{2}\I(D^\al w_\al + \bar D_\da \bar
w^\da) \Gamma = -\tfr{3}{2} \I (D^2 S^{\rm conf} + \bar D^2 \bar S^{\rm
  conf}) \cdot \Gamma
\end{align}
(which incidentally explains the origin of the name ``trace equation'' for (\ref{traceeq})).

It is now crucial to observe that the hard terms in $S^{\rm conf}$ can be
represented by a local functional operator which commutes with $\tilde w^3$
\begin{equation}
\Bl \equiv \half \sum_{r=1}^\infty \gamma^{(r)} \hbar^r \bar D^2 \left\{
  (\Lambda \bar   \Lambda)^r \e^{-3rL} \fdq{}{L} \right\}\,,
\end{equation}
acting on $\Gamma$, i.e.\
\begin{equation}
\left( w_{\rm trace}^P + \I D^2 \Bl+ \I \bar D^2 \Blb \right) \Gamma =
\tfr{\I}{8} m [D^2 A^2 + \bar D^2 \bar A^2]_3 \cdot
\Gamma \label{CSlocal}
\end{equation}

In order to prove this we first calculate
\begin{align}
\Bl \Geff &= \phantom{+}\sum_{n,m} \sum_{r=1}^n \sum_{s=0}^m \tfr{1}{32}
\frac{(-3r)^s}{s!}\gamma^{(r)} z^{(n-r, m+1-s)} (m+1-s) \hbar^n \Lkin^{(n,m)} \nonumber \\
&\quad + \sum_{n,m} \sum_{r=1}^n \sum_{s=0}^m \tfr{1}{16} \frac{(-3r)^s}{s!}\gamma^{(r)}
\xi^{(n-r, m+1-s)} (m+1-s) \hbar^n (\Lxi^{(n,m)} + \Lxib^{(n,m)})
\end{align}
This coincides with the hard terms in $S^{\rm conf}$ if
\begin{align}
0  &= \ukin^{(n,m)} + \quar \sum_{r=1}^n \sum_{s=0}^m \frac{(-3r)^s}{s!}\gamma^{(r)} z^{(n-r,
  m+1-s)} (m+1-s) & &\equiv q_1(n,m) \label{CSeq1}\\
0 &=\uxi^{(n,m)} + \uxib^{(n,m)} + \sum_{r=1}^n \sum_{s=0}^m
  \frac{(-3r)^s}{s!}\gamma^{(r)}\xi^{(n-r,  m+1-s)} (m+1-s)  && \equiv q_2(n,m)  \,.
  \label{CSeq2}
\end{align}
By properly adjusting $\gamma^{(r)}$, $q_1(n,0)$ can be made to vanish,
 \begin{align}
 \gamma^{(1)}&= -4 \ukin^{(1,0)} \\
 \gamma^{(n)} &= -4\ukin^{(n,0)} -\sum_{r=1}^{n-1} \gamma^{(r)}
 z^{(n-r,1)} & (n\ge 2)
 \end{align}

Before we show that the remaining $q_1$, $q_2$ vanish automatically, we
pass over from (\ref{CSlocal}) to the integrated CS equation.
\begin{equation}
W^D \Gamma +\I \B \Gamma = \tfr{\I}{8} m \left[ \int \dS A^2 + \int \dSb
  \bar A^2 \right]_3 \cdot\Gamma\,,
\end{equation}
where $\B$ is the integrated version of $\Bl$,
\begin{equation}
\B = \int \dS \Bl + \int \dSb  \Blb\,.
\end{equation}
Thus the integrated CS equation reads
\begin{equation}
\C \Gamma =\tfr{1}{8} m \left[ \int \dS A^2 + \int \dSb \bar A^2
\right]_3 \cdot\Gamma \label{CSint}
\end{equation}
with the CS operator
\begin{equation}
\C =  m\pr_m + 2 \kappa^2 \pr_{\kappa^2} +\B \,.
\end{equation}
In order to finally prove (\ref{CSeq2}),
we consider the commutator
\begin{equation}
\left[\C,  \tilde w^3 (z)\right]=0\,.
\end{equation}
Since $\tilde w^3\Gamma=\text{soft}$, this implies
\begin{equation}
 \tilde w^3(z) \, \C \Gamma=\text{soft} \,.
\end{equation}
For the lowest non-vanishing order $n$ this means
\begin{align}
q_1(n,m+1) &= \frac{3n-1}{m+1} q_1(n,m)\\
q_2(n,m) &= 0\,.
\end{align}
Since $q_1(n,0)=0$ due to our choice of $\gamma^{(r)}$,
this shows that (\ref{CSeq1}) and (\ref{CSeq2}) hold, and we have thus
established the local and integrated CS equations (\ref{CSlocal}),
(\ref{CSint}). Moreover, the
trace equation (\ref{traceeq}) may now be nicely expressed as
\begin{equation}
-2 w_\al^{\rm conf} = \bar D^\da V^{\rm conf}_{\al\da} \cdot\Gamma+ \tfr{4}{3} D_\al \Bl \Gamma
 -\tfr{1}{6} m [D_\al A^2]_{5/2}\cdot\Gamma\,.
\label{traceequation2}
\end{equation}

In order to make contact with the standard Wess-Zumino model, we take
the limit $L=0$, $\Lambda=\bar\Lambda=\lambda=const$. First we define
\begin{align}
z &= \sum_n z^{(n,0)} \hbar^n \lambda^{2n} & 
\xi &= \sum_n \xi^{(n,0)} \hbar^{n+1} \lambda^{2n+2} \\
\gamma &= -\half \sum_r \gamma^{(r)} \hbar^r \lambda^{2r}
 &
\beta &= 3\lambda \gamma \label{betagamma} \\
\tilde z &= \sum_n z^{(n,1)} \hbar^n \lambda^{2n} & 
\ukin &= \sum_n \ukin^{(n,0)} \hbar^n \lambda^{2n}
\end{align}
From (\ref{CSeq1}), (\ref{zsolv}) we know that
\begin{align}
\gamma &= \frac{2\ukin}{\tilde z}\,, &
\tilde z &= (1-\tfr{3}{2} \lambda\pr_\lambda) z + 4\ukin\,,
\end{align}
such that the standard form for $\gamma$ is recovered,
\begin{equation}
\gamma = \frac{2\ukin}{z+4\ukin -\tfr{3}{2} \lambda \pr_\lambda z}\,.
\end{equation}
In the considered limit, the operator $\Bl$ takes a simple form,
\begin{equation}
\Bl \to -\gamma \bar D^2 \fdq{}{L}\,.
\end{equation}
Taking into account the gauge WI (\ref{gaugeinv}), we find
\begin{align}
\Bl \Gamma  & \to -\gamma A \fdq{}{A} \Gamma + \beta \fdq{}{\Lambda} \Gamma +
\quar \gamma m [A^2]_2\cdot\Gamma \\
\B \Gamma & \to \left( -\gamma \N_A +
 \beta   \pr_\lambda\right) \Gamma +\quar \gamma  m
  \left[ \int \dS A^2 + 
  \int\dS \bar A^2\right]_3\cdot\Gamma\,, 
\end{align}
which reproduces the usual CS equation (\ref{CSstandard}). The crucial
point is that in the Wess-Zumino model with local coupling a
Callan-Symanzik equation can be formulated only with the help of an
additional external field, but in the limit of constant coupling this field
couples just to the coupling constant and wave function renormalization
operators. 
In our approach, the relation (\ref{betagamma}) between the $\beta$ and
$\gamma$ functions is a consequence of the gauge WI (\ref{gaugeinv}).
 
It is furthermore interesting to note that $\fdq{}{\Lambda}\Geff$ may be
interpreted as the $\lambda$-derivative of a Lagrangian density,
\begin{align}
\left.\fdq{\Geff}{\Lambda}\right|_{L=0 \atop \Lambda=\lambda} &=
\pr_\lambda \Leff\\ 
\left.\Geff\right|_{L=0 \atop \Lambda=\lambda} &= \int \dS \Leff + \int
\dSb \Leffb \\
\Leff &\equiv \tfr{1}{32} z \bar D^2 (A\bar A) + \tfr{1}{48} \lambda A^3 
+ \tfr{1}{8} \eps \bar D^2 \bar A^2 \\
\eps &\equiv \sum_n \xi^{(n,0)} \hbar^{n+1} \frac{n}{2n+2} \lambda^{2n+2}\,,
\end{align}
such that (\ref{traceequation2}) reduces to
\begin{equation}
-2 w_\al^{(\gamma)} \Gamma = \bar D^\da   V_{\al\da} \cdot \Gamma
+ \tfr{4}{3} D_\al ( \beta \pr_\lambda \Leff)  \cdot \Gamma 
-\tfr{1}{6} (1-2\gamma) m D_\al [A^2]_2\cdot\Gamma
\end{equation}
with the anomalous contact terms
\begin{equation}
w_\al^{(\gamma)} \equiv w_\al^{\rm conf} -\tfr{2}{3} \gamma D_\al(A
\fdq{}{A})
\,.
\end{equation}
This result coincides with the flat space limit of the construction of
\cite{ERS1}. 

\section{Application}
\setcounter{equation}{0}

The main reason to introduce a local coupling constant is that it gives
some new insight into the non-renormalization theorems of chiral vertices
\cite{FK}. The present paper extends the considerations of \cite{FK} in
supplying the Callan-Symanzik equations established in the previous
section and thus completing the proof of renormalizability for the model
under consideration.
However, the external fields $\Lambda$ and $L$ may also be used to generate
insertions of local operators,
\begin{align}
\Delta_L &\equiv \fdq{\Geff}{L} \Bigr|_{L=0,\Lambda=\lambda} = \tfr{1}{16} A \bar A + O(\hbar)\\
\Delta_\lambda &\equiv \fdq{\Geff}{\Lambda} \Bigr|_{L=0,\Lambda=\lambda} = \tfr{1}{48} A^3 + O(\hbar)\,.
\end{align}
Since we are working with off-shell Ward
identities, we also have access to the covariance properties of these
operator insertions. 
For example, we find for the dilatational transformation of the
$\Delta_L$-insertion: 
\begin{align}
W^D \left( [\Delta_L]\cdot \Gamma\right) &\sim [\delta^D \Delta_L]\cdot \Gamma
+ (-\gamma \N_A + \beta \pr_\lambda + 3 \lambda \pr_\lambda \gamma)
([\Delta_L]\cdot\Gamma) \label{DeltaLtrans}\\
\intertext{with}
\delta^D \Delta_L &= (x^\mu\pr_\mu + \half \theta^\al \pr_\al -\half
\bar\theta_\da \bar\pr^\da +2) \Delta_L \label{DeltaLtransD}
\end{align}
Here, $W^D$ represents only the transformation of the fields $A$, $\bar
A$. $\sim$ means equality up to soft terms. 
Since $\Delta_L$ contains two dynamical fields, one might naively expect
its anomalous dimension to be $2\gamma$. Instead we find that its
anomalous dimension is given by $ 3 \lambda \pr_\lambda \gamma$.

The scaling behavior of $\Delta_\lambda$ is given by
\begin{align}
W^D  \left( [\Delta_\lambda ]\cdot \Gamma\right) &\sim  [\delta^D
\Delta_\lambda]\cdot \Gamma + (-\gamma \N_A + \beta \pr_\lambda + 3\gamma)
([\Delta_\lambda]\cdot\Gamma)  - \lambda\pr_\lambda \gamma [\bar D^2
\Delta_L]\cdot\Gamma  \\
\delta^D \Delta_\lambda &= (x^\mu\pr_\mu + \half \theta^\al \pr_\al -\half
\bar\theta_\da \bar\pr^\da +3) \Delta_\lambda
\end{align}
Hence this operator indeed has the naive anomalous dimension $3\gamma$, but
dilatations do not close on $\Delta_\lambda$ alone: the
operator $\Delta_L$ is also involved here.

One can also produce multiple insertions of $\Delta_\lambda$,
$\Delta_L$. For example, it follows from (\ref{gaugeinv}) by differentiating
w.r.t.\ $L$ and then taking the adiabatic limit that
\begin{equation}
\lambda  \l \{ \Delta_\lambda(z_1) \cdot \Delta_L(z_2)\}\r \sim
-\tfr{1}{3}  \l \{
\bar D^2 \Delta_L(z_1) \cdot \Delta_L(z_2)\}\r + u_{\rm geom} \bar
D^2 \Box \delta^8(z_1-z_2)\,.
\end{equation}
The last term represents the only possible contribution from $\Delta_{\rm
  triv}$ (\ref{Zimmermann}). $u_{\rm geom}$ is a power series in $\lambda$ and
$\hbar$. For the definition of double insertions, see \cite{ER}.
Differentiation of (\ref{gaugeinv}) w.r.t.\ $\Lambda$ yields
\begin{equation}
\lambda \l \{ \Delta_\lambda(z_1)\cdot \Delta_\lambda(z_2)\} \r \sim
-\tfr{1}{3} \l \{ \bar D^2 \Delta_L(z_1)\cdot \Delta_\lambda(z_2)\}\r\,,
\end{equation}
this time $\Delta_{\rm triv}$ does not contribute.

\section{Conclusions}
\setcounter{equation}{0}
We have constructed the Wess-Zumino model with local coupling represented
by a chiral external field ${\Lambda}$ and derived
the supercurrent in its presence. A further necessary ingredient was an external
real superfield $L$ coupled to the composite operator $A\bar{A}$. From a
merely technical point of view it is this composite operator which causes
all renormalizations in the model once one has encoded the fact that
chiral vertices are not renormalized. The wavefunction renormalization
and subsequently the anomalous dimension of the chiral field $A$ as well as
the $\beta$-function of the model can all be related to the effect of inserting
$A\bar A$: certainly the most clear cut implementation of the
non-renormalization of chiral vertices and its consequences. Speaking in
equations the consequences are expressed by the local CS equation which
was derived in the context
of the supercurrent with its moment construction giving access to the
entire superconformal group. Like in the other examples of the local
coupling approach it is an axial current  and its WI which plays the most
decisive role in the derivation of these results: the respective local
WI leads one to the interrelations of all potential anomaly coefficients.
The local coupling and the external field $L$ also serve the purpose
of defining multiple insertions of the respective composite operators
$A^3$ and $A\bar A$ including their transformations under the superconformal
group. It is also interesting to observe that the local coupling approach
yields some results which had been obtained previously by going to the curved
superspace. Hence there is a connection which deserves further study.
This comment also applies to the study of multiple insertions to which we hope to
come back in the near future. Summarizing we may say that the introduction
of a local coupling is without any doubt the key to understand all
renormalization questions in the Wess-Zumino model.

\def\npb#1#2#3{{Nucl. Phys. }{\bf B#1\ }(19#2)\ #3}
\def\npbr#1#2#3{{Nucl. Phys. }{\bf B#1\ }(20#2)\ #3}
\def\prd#1#2#3{{Phys. Rev. }{\bf D#1~}(19#2)~#3}
\def\plb#1#2#3{{Phys. Lett. }{\bf B#1~}(19#2)~#3}

\end{document}